\documentclass[11pt,letterpaper]{article}
\usepackage[margin=0.75in,top=0.67in,bottom=0.66in,headheight=14pt,headsep=10pt,footskip=23pt]{geometry}
\usepackage[T1]{fontenc}
\usepackage{newtxtext}
\usepackage{amsmath,amsthm}
\usepackage{newtxmath}
\ifdefined\pdfglyphtounicode
  \pdfglyphtounicode{summationtext}{2211}
  \pdfglyphtounicode{summationtext.1}{2211}
  \pdfglyphtounicode{parenleftbigg}{0028}
  \pdfglyphtounicode{parenrightbigg}{0029}
\fi
\usepackage{microtype,booktabs,tabularx,array,caption,fancyhdr,titlesec,tikz}
\usetikzlibrary{arrows.meta,decorations.pathreplacing}
\usepackage{hyperref}
\definecolor{ink}{HTML}{20364A}
\definecolor{teal}{HTML}{006D70}
\definecolor{ochre}{HTML}{995323}
\hypersetup{colorlinks=true,linkcolor=ink,citecolor=teal,urlcolor=teal,
  pdfauthor={C. S. Elder},
  pdftitle={Square-root comparison, Turnpike, and exact semidefinite feasibility},
  pdfsubject={Exact conic feasibility and the representation of unlabeled distances}}
\fancypagestyle{firstpage}{\fancyhf{}\fancyfoot[C]{\small\thepage}}
\titleformat{\section}{\large\bfseries\color{ink}}{\thesection}{0.55em}{}
\titleformat{\subsection}{\normalsize\bfseries\color{ink}}{\thesubsection}{0.55em}{}
\titlespacing*{\section}{0pt}{0.8\baselineskip}{0.32\baselineskip}
\titlespacing*{\subsection}{0pt}{0.64\baselineskip}{0.24\baselineskip}
\theoremstyle{plain}
\newtheorem{proposition}{Proposition}
\newtheorem{theorem}[proposition]{Theorem}
\newtheorem{corollary}[proposition]{Corollary}
\DeclareMathOperator{\sep}{sep}
\DeclareMathOperator{\gap}{gap}
\DeclareMathOperator{\dist}{dist}
\newcommand{\R}{\mathbb R}
\newcommand{\Q}{\mathbb Q}
\newcommand{\Z}{\mathbb Z}
\newcommand{\psd}{\succeq 0}
\newcommand{\summaryhead}[1]{\par\smallskip\noindent\textbf{#1}\enspace}
\AtBeginDocument{%
  \setlength{\abovedisplayskip}{7pt plus 2pt minus 2pt}%
  \setlength{\belowdisplayskip}{7pt plus 2pt minus 2pt}%
  \setlength{\abovedisplayshortskip}{3pt plus 1pt}%
  \setlength{\belowdisplayshortskip}{4pt plus 1pt minus 1pt}%
  \setlength{\topsep}{4pt plus 1pt minus 1pt}%
}

\title{Square-root comparison, Turnpike,\\and exact semidefinite feasibility}
\author{C.~S.~Elder}
\date{25 November 2026}
\makeatletter
\renewcommand{\maketitle}{%
  \noindent{\LARGE\bfseries\color{ink}\@title\par}
  \vspace{0.4\baselineskip}
  \noindent\@author\hfill{\small\@date}\par
  \vspace{0.6\baselineskip}}
\makeatother
\begin{document}
\thispagestyle{firstpage}
\maketitle
\noindent\textbf{Abstract.}
Given positive integers \(a_i,K\), put \(S=\sum_{i=1}^k\sqrt{a_i}\). We construct
a realizable Turnpike instance in which the minimum spacing of distinct distances
and the smallest nonzero addition residual are both at most \(|S-K|\) when
\(S\ne K\). With approximation oracles, one three-distance realizability query
decides \(S\le K\). For unordered lists of radical sums with nonnegative rational
coefficients, a polynomial-time substitution by integers preserves every normalized
ruler. We give the classical reduction of square-root comparison to exact
semidefinite feasibility in both threshold directions. The comparison enters
through numerical separation and marked sorting; the integer recoding needs
no such comparison for unordered reconstruction.

\section{Two small gaps on a realizable ruler}\label{sec:gaps}
Deciding \(S\le K\) in time polynomial in the binary input length \(L\)
is a longstanding open problem~\cite[p.~658]{goemans_sdp_1998}.
The exact Turnpike problem asks whether a multiset \(Y\) of positive lengths equals
\(\dist(X)=\{x_j-x_i:i<j\}\), with multiplicity, for some \(x_1<\cdots<x_n\).
Write \(D\) for the distinct labels and \(\mu(u)\) for the multiplicity of \(u\).
A \emph{two-partition} is an equality \(u+v=w\) in \(D\), with
\(\mu(u)\ge2\) if \(u=v\). The triangle-equality formulation uses these
relations to assign distances to intervals
\cite[Prop.~10]{elder_turnpike_with_uncertain_measurements_wabi_2026}.
The two numerical margins are
\begin{equation}
 \sep(Y)=\min_{\substack{u,v\in D\\u\ne v}}|u-v|,
 \qquad
 \gap_\star(Y)=\min_{\substack{u,v,w\in D\\u+v\ne w}}|u+v-w|.
 \label{eq:gaps}
\end{equation}
An empty minimum is \(+\infty\). The second minimum ranges over all numerical
triples, including repeated labels, as in the published definition
\cite[Thm.~15]{elder_turnpike_with_uncertain_measurements_wabi_2026}.

\begin{proposition}\label{prop:planted}
Let \(k\ge1\), \(K>0\), and \(\Delta=S-K\). In polynomial bit time one can
construct the complete distance multiset \(Y\) of a \((k+3)\)-site ruler as
explicit sums of radicals. Whenever \(\Delta\ne0\),
\(\sep(Y)\le|\Delta|\) and \(\gap_\star(Y)\le|\Delta|\).
\end{proposition}
\begin{proof}
Set \(B=2(1+K+\sum_i a_i)\). Use consecutive gaps
\(\sqrt{a_1},\ldots,\sqrt{a_k},K,B\), as in Figure~\ref{fig:ruler}.
The four lengths \(S,K,B,K+B\) occur as intervals. Since
\(\sqrt{a_i}\le a_i\), we have \(B>\max(S,K)\). For \(\Delta\ne0\), the
labels \(S,K\) are distinct. The extra gap \(B\) gives a nonzero residual
\(S+B-(K+B)=\Delta\) on three distinct labels. These witnesses
give the two upper bounds in~\eqref{eq:gaps}. Listing all contiguous gap sums
takes \(O(k^2)\) intervals with at most \(k+2\) terms apiece. The integers
in each term have polynomial bit length.
\end{proof}

\par\smallskip\begingroup\centering\noindent
\begin{minipage}{0.94\linewidth}
\centering
\begin{tikzpicture}[x=0.92cm,y=0.81cm,>=Stealth,
  brace/.style={decorate,decoration={brace,amplitude=3.6pt},semithick}]
  \draw[very thick,ink] (0,0)--(12.25,0);
  \foreach \x in {0,4.8,7.25,12.25} \fill[ink] (\x,0) circle (2.2pt);
  \draw[teal,brace] (0,.19)--(4.8,.19) node[midway,above=5pt] {\(\sqrt{a_1}+\cdots+\sqrt{a_k}=S\)};
  \draw[ink,brace] (4.8,.19)--(7.25,.19) node[midway,above=5pt] {\(K\)};
  \draw[ink,brace] (7.25,.19)--(12.25,.19) node[midway,above=5pt] {\(B\)};
  \node[below=3pt] at (0,0) {\(0\)};
  \node[below=3pt] at (4.8,0) {\(S\)};
  \node[below=3pt] at (7.25,0) {\(S+K\)};
  \node[below=3pt] at (12.25,0) {\(S+K+B\)};
  \draw[ochre,brace] (12.25,-.73)--(4.8,-.73) node[midway,below=5pt] {\(K+B\)};
\end{tikzpicture}

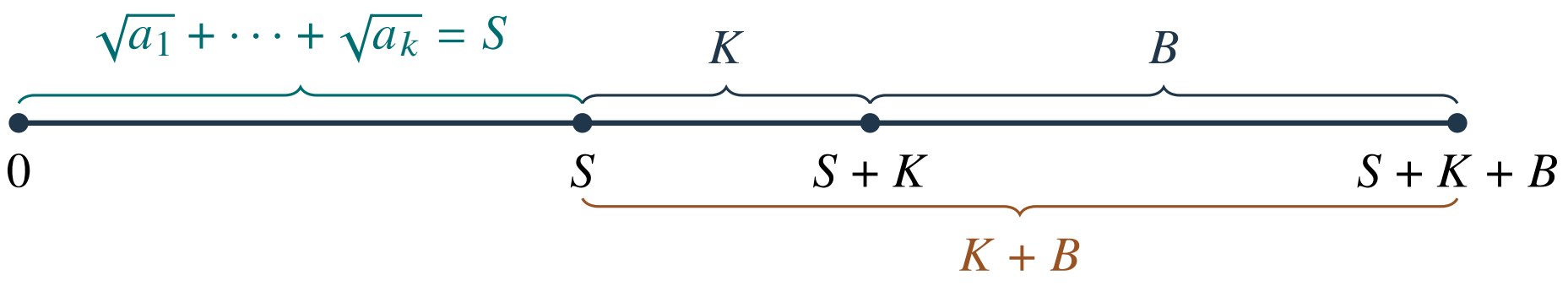
\captionof{figure}{The first span exposes \(S\), and the tail contributes
\(K,B,K+B\). Substituting \(S\) for \(K\) in the tail's true addition gives
the residual \(S-K\). Interior sites in the first span are omitted. Lengths are schematic.}\label{fig:ruler}
\end{minipage}
\par\endgroup\smallskip

\summaryhead{Why the two margins matter.}
Given one representative per true label with error at most \(\eta\), any
residual changes by at most \(3\eta\). Thus \(\gap_\star>6\eta\) makes
\(|\widehat u+\widehat v-\widehat w|\le3\eta\) an exact equality test,
and \(\sep>2\eta\) preserves label order
\cite[Ch.~4]{elder_robust_unlabeled_distance_recovery_2026}.
For separately perturbed occurrences, \(\sep>4\eta\) also recovers
multiplicities by cutting the sorted observations at gaps larger than
\(2\eta\). Each class spans at most \(2\eta\), and consecutive classes are
farther apart.

On a ruler with at least three sites, \(\gap_\star\le\sep\). Take two
closest distinct labels. Any occurrence of the first belongs to a triangle
with a third site. Replacing one copy in that triangle equality by the
second label gives a residual of magnitude \(\sep\).
The ratio \(\sep/\gap_\star\) can be arbitrarily large. The ruler
\((0,1,4,9+t)\), \(0<t<1\), has distances \(1,3,4,5+t,8+t,9+t\).
Their minimum spacing is \(1\), while \(1+4-(5+t)=-t\).

\begin{corollary}\label{cor:precision}
An explicit uniform lower bound \(2^{-p(L)}\), for a known polynomial \(p\),
on either margin in Proposition~\ref{prop:planted} for \(\Delta\ne0\)
would put the positive-summand threshold problem in polynomial bit time.
The same implication holds for a bound stated in terms of the length of
the constructed list.
\end{corollary}
\begin{proof}
Square roots of distinct square-free positive integers, including \(1\),
are linearly independent over \(\Q\)
\cite[Thm.~1]{eisenbrand_haeberle_singer_2024}.
After grouping by square-free part, every nonsquare class present in \(S\)
has positive coefficient. Thus \(S=K\) can hold only if every \(a_i\) is
a square. Integer square-root tests and, in that case, addition decide equality
without factoring. For \(\Delta\ne0\) the proposed margin bound gives
\(|\Delta|\ge2^{-p(L)}\). Approximating each root to absolute error less
than \(2^{-p(L)-1}/k\) determines the sign of their sum minus \(K\).
Dyadic integer square roots achieve this precision in polynomial time.
Proposition~\ref{prop:planted} bounds the length of the constructed list
polynomially in \(L\). The case \(K\le0\) is immediate since \(S>0\).
\end{proof}
Actual near ties transfer directly to rulers. For fixed \(k\ge2\) and large
\(N\), Steinerberger gives \(a_i\le N\) and an integer \(K\) with
\(0<|S-K|\le C_k N^{-c k^{1/3}}\) for an absolute \(c>0\)
\cite{steinerberger_close_2024}. This gives an upper bound on how small the
two margins can be. For signed sums, the lower bound of Eisenbrand, Haeberle,
and Singer has a nonexplicit radicand-dependent constant
\cite{eisenbrand_haeberle_singer_2024}. A uniform polynomial-bit guarantee
for variable radicands remains a separate question.

\section{What the Turnpike input reveals}\label{sec:interfaces}
\subsection{Marked order and approximation names}
For either sign of \(\Delta\), the planted ruler exists. Mark the interval \(S\)
and the gap \(K\). After testing equality, their positions in a sorting permutation
give the sign. The earlier Turnpike optimization algorithm uses such order in
its sorted matching~\cite[Prop.~1]{elder_approximate_and_exact_optimization_algorithms_2024}.

\begin{proposition}\label{prop:name}
One exact Turnpike query on three distances presented by approximation
names decides \(S\le K\). The names are uniformly computable in time
polynomial in the binary input length and the requested precision.
\end{proposition}
\begin{proof}
A name for \(x\) returns a dyadic within \(2^{-p}\) of \(x\) when asked for
\(p\) bits after the binary point. Set \(t_+=\max(0,t)\) and query
\[
                  \{1,\ 2,\ 3+(\Delta)_+\}.
\]
Three positive distances form a ruler precisely when the largest equals
the sum of the other two. Indeed the two gaps of a three-site ruler determine
its span, and any two positive gaps give such a ruler. The query is
therefore feasible exactly when \(\Delta\le0\).
For the third name, use \(q=p+\lceil\log_2 k\rceil+2\) and approximate
\(\sqrt{a_i}\) by \(2^{-q}\lfloor\sqrt{2^{2q}a_i}\rfloor\).
The sum has absolute error at most \(k2^{-q}\), and taking the positive
part is \(1\)-Lipschitz. Integer arithmetic on \(O(L+q)\)-bit inputs
produces the answer in polynomial time without testing the sign of \(\Delta\).
\end{proof}
Exact realizability is discontinuous on arbitrary names. Run a putative decider
on constant answers \(1,2,3\), and let \(p_*\) be the greatest precision it queries
before halting (zero if none). For dyadic \(0<\delta<2^{-p_*}\), every third-coordinate
reply also fits \(3+\delta\). The same transcript would give the wrong answer.
For finite expressions admitting a positive-part gate, the displayed formula
gives a polynomial-size instance directly.

\par\smallskip\begingroup\centering\noindent
\begin{minipage}{0.98\linewidth}
\captionof{table}{Which Turnpike operation carries the comparison?}\label{tab:interfaces}
\small\renewcommand{\arraystretch}{1.13}
\begin{tabularx}{\linewidth}{@{}l l >{\raggedright\arraybackslash}X@{}}
\toprule
Distance input & Requested output & What follows\\
\midrule
Marked intervals in a valid ruler & Sorting permutation & The order of \(S,K\) gives the sign.\\
Three approximation names & Exact realizability & One oracle query decides \(S\le K\).\\
Unordered radical lists & Exact realizability & Equivalent to integer Turnpike (Theorem~\ref{thm:integer}).\\
\bottomrule
\end{tabularx}
\end{minipage}
\par\endgroup\smallskip

\subsection{Unordered lists of explicit radical sums}
Suppose each positive distance is supplied as a list of terms \(c\sqrt a\)
with \(a\) a positive binary integer and \(c\) a nonnegative binary-encoded
rational. The input consists of the distance occurrences in arbitrary order.
Normalize a ruler by translating its leftmost site to zero.

\begin{theorem}\label{thm:integer}
Turnpike on these radical lists and Turnpike on binary integers are
polynomial-time many-one equivalent. In the forward direction, a dictionary
of distance labels induces a bijection between all normalized realizations.
\end{theorem}
\begin{proof}
Discard zero terms. Two radicands \(a,b\) belong to the same rational
square class exactly when \(ab\) is a perfect square, since \(a/b=ab/b^2\).
For a representative \(b\) of that class,
\(\sqrt a=(\sqrt{ab}/b)\sqrt b\). Pairwise perfect-square tests therefore
produce, for each label \(u\), a nonnegative rational coefficient vector on
one representative per class. The representatives are rational multiples of
distinct square-free radicals, so independence makes vector equality
equivalent to equality of the labels. Merge equal labels, retaining their
multiplicities.

We use the classical large-base construction of a Freiman isomorphism
\cite[Lemma~5.25]{tao_vu_additive_2006}.
Clear the denominators of all vectors by a common positive multiplier.
Write \(w(u)\in\Z_{\ge0}^{r}\) for the resulting vector and let \(M\) be its
largest entry over all labels. For a nonempty positive input \(M\ge1\).
Set \(\beta=2M+1\) and encode each label by
\begin{equation}
              f(u)=\sum_{j=1}^{r}w_j(u)\,\beta^{j-1}.
              \label{eq:digits}
\end{equation}
Each label maps to a positive integer. The sum of any two coordinate digits
is at most \(2M<\beta\), so addition of two encoded labels creates no carry.
Uniqueness of base-\(\beta\) digits gives injectivity and, in both directions,
\[
                      u+v=w \quad\Longleftrightarrow\quad f(u)+f(v)=f(w).
\]
All arithmetic so far is polynomial in the input length. In particular the
product of the converted term denominators has bit length at most the sum
of their bit lengths. There are polynomially many such terms, each obtained
by integer square root and rational arithmetic on polynomial-bit inputs.
Thus \(\log\beta\) is polynomially bounded, as is \(r\), and each output
integer is less than \(\beta^r\).

Every positive site of a normalized realization is a distance from zero, so
it has a dictionary entry. Extend \(f\) by \(f(0)=0\). For positive sites
\(x<y\), let \(z=y-x\). Then \(z\) is an input label and \(x+z=y\), hence
\(f(y)-f(x)=f(z)>0\). Distances from zero transform by definition.
Consequently every pair of sites realizes the prescribed image label with
its multiplicity. Conversely, in a normalized integer realization every
positive site and every pairwise difference occurs in the image dictionary.
Applying \(f^{-1}\) to their addition identities gives positive differences
in the original labels. The two substitutions are mutual inverses.
Integer inputs can be written as \(u\sqrt1\). The empty distance list,
if admitted for a one-site ruler, maps to itself.
\end{proof}

For example, the labels \(\sqrt2,3,3+\sqrt2\), using representatives
\(1,\sqrt2\) in that order, map to \(7,3,10\). Although \(\sqrt2<3\), the
encoded value \(7\) exceeds \(3\). The normalized ruler \((0,\sqrt2,3+\sqrt2)\)
maps to \((0,7,10)\), and its reflection maps to \((0,3,10)\).
The radical syntax supplies the equalities. The proof needs
the order of sites that coexist in a realization, which their positive
differences already enforce.

\summaryhead{Decision complexity.}
Ordinary integer Turnpike is in \(\mathrm{NP}\) for an explicit distance list.
A certificate gives the normalized sites, each zero or an input distance.
One checks their number, their strict order, and all pairwise differences
with multiplicity in polynomial time. Theorem~\ref{thm:integer} gives the
same upper bound for radical lists. A reduction from \(S\le K\) to either
unordered decision problem would therefore place the threshold problem in
\(\mathrm{NP}\). The marked sort and the query on names achieve their
comparisons through the different operations in Table~\ref{tab:interfaces}.

\section{Which conic feasibility algorithm would suffice?}\label{sec:sdp}
A rational-input semidefinite feasibility problem asks whether a rational
affine symmetric matrix pencil is positive semidefinite at some
\emph{real} choice of its variables. Blocks are assembled by direct sum,
with a scalar block for a linear inequality. Goemans identified square-root
comparison as a special case~\cite[p.~658]{goemans_sdp_1998}.
Writing both threshold directions explicitly matters because the familiar
block for \(|x|\le\sqrt a\) puts an \emph{upper} bound on \(x\).

\begin{proposition}\label{prop:sdp}
For \(a>0\) set
\[
 L_a(x)=\begin{pmatrix}x&a\\a&ax\end{pmatrix},\qquad
 U_a(x)=\begin{pmatrix}a&x\\x&1\end{pmatrix}.
\]
Both threshold decisions reduce in polynomial bit time to feasibility of
rational affine matrix pencils.
\begin{align}
 S\le K &\quad\Longleftrightarrow\quad
    \exists x\in\R^k:\quad L_{a_i}(x_i)\psd\ \ (\forall i),\quad \textstyle\sum_i x_i\le K ,
       \label{eq:lower}\\
 S\ge K &\quad\Longleftrightarrow\quad
    \exists x\in\R^k:\quad U_{a_i}(x_i)\psd\ \ (\forall i),\quad \textstyle\sum_i x_i\ge K .
       \label{eq:upper}
\end{align}
\end{proposition}
\begin{proof}
The diagonal and determinant tests for a symmetric \(2\times2\) matrix give
\(L_a(x)\psd\) exactly when \(x\ge0\) and \(a(x^2-a)\ge0\), or \(x\ge\sqrt a\).
Similarly \(U_a(x)\psd\) exactly when \(a-x^2\ge0\), or \(|x|\le\sqrt a\).
The smallest attainable sum in~\eqref{eq:lower} before the linear cut is
\(S\), and the largest in~\eqref{eq:upper} is \(S\). Both are attained at
\(x_i=\sqrt{a_i}\). Including the scalar cut gives one block-diagonal pencil
of order \(2k+1\), with polynomial encoding length in the binary data.
\end{proof}
The maximum slack of the cut in~\eqref{eq:lower} is \(K-S\). If \(K>S\), setting \(x_i=\sqrt{a_i}+(K-S)/(2k)\)
gives strict feasibility. Equality forces every endpoint. An approximation
certifies the sign if its error interval excludes \(K\).

For integer \(K\), equality requires square radicands
(Corollary~\ref{cor:precision}). In general the witness can be irrational:
\(L_2(x)\psd\) and \(U_2(x)\psd\) together force \(x=\sqrt2\). Exact SDP
uses real witnesses~\cite{ramana_exact_duality_1997}.

\subsection{Why the new LP algorithm does not settle this comparison}
OpenAI's preprint reports a strongly polynomial algorithm for explicit LP.
Its Corollary~6 minimizes a convex quadratic objective over a polyhedron
\cite[p.~36]{openai_strong_lp_2026}. Strong polynomiality bounds the number of
arithmetic operations and comparisons by a polynomial in the dimensions.
On rational input it also bounds intermediate bit lengths polynomially in the
full input length. Writing \(x_i\ge\sqrt{a_i}\) as a linear constraint
would put the very radical being compared into the LP data.

Corollary~7 of the same preprint treats a single Lorentz constraint
\(\|F(x)\|_2\le h(x)\) with affine \(F,h\), and multiple such constraints
when the Hessians of \(\|F\|_2^2-h^2\) span a space of dimension at most one
\cite[p.~37]{openai_strong_lp_2026}.
This hypothesis can be checked directly for Proposition~\ref{prop:sdp}.
For a real symmetric block the trace and determinant tests give
\[
 \begin{pmatrix}u&v\\v&w\end{pmatrix}\psd
 \quad\Longleftrightarrow\quad \|(u-w,2v)\|_2\le u+w .
\]
The squared residual (left side squared minus right side squared) is
\(4(v^2-uw)\). For \(U_{a_i}(x_i)\) it is \(4(x_i^2-a_i)\), and for
\(L_{a_i}(x_i)\) it is \(4a_i(a_i-x_i^2)\). In the common variable
\(x\in\R^k\) the respective Hessians are
\begin{equation}
       8e_ie_i^{\mathsf T}\qquad\text{and}\qquad
       -8a_i e_ie_i^{\mathsf T}.
       \label{eq:hess}
\end{equation}
Here \(e_i\) is the \(i\)th coordinate vector. The disjoint diagonal supports
give span dimension \(k\) in each family. Thus Corollary~7 applies directly
when \(k=1\). For \(k>1\) its hypothesis fails for these encodings.
Reformulation can change the span: \(\sqrt2+\sqrt8=3\sqrt2\) needs only one block.

\summaryhead{Consequences for Turnpike.}
An exact polynomial-bit-time solver for~\eqref{eq:lower} would decide \(S\le K\)
and order the marked lengths of Proposition~\ref{prop:planted}. The displayed
family also determines \(a_1,\ldots,a_k,K\) by inspection. For unrestricted SDP
only the reduction from comparison to feasibility has been established here.

To compare two sums with nonnegative rational coefficients \(c_i,d_j\), put
lower blocks on the left, upper blocks on the right, and impose
\(\sum_i c_i x_i\le\sum_j d_j y_j\). The endpoints give equivalence without
computing a numerical gap. For unordered Turnpike, Theorem~\ref{thm:integer}
leaves the assignment of labels to intervals. The triangle ILP enforces
this assignment from the equality table. Tardos's algorithm applies to
its small-integer LP relaxation. A ruler requires an integral assignment
\cite{elder_turnpike_with_uncertain_measurements_wabi_2026,tardos_a_strongly_polynomial_algorithm_to_solve_1986}.

\smallskip\noindent{\small\textbf{Acknowledgment.} OpenAI's ChatGPT and Codex assisted with proof development and checking, reference searches, figure preparation, and manuscript revision.}

\end{document}